# Particle and thermo-hydraulic maldistribution of nanofluids in parallel microchannel systems

**Lakshmi Sirisha Maganti [a], Purbarun Dhar [b], T. Sundararajan\*, [c] and Sarit K. Das\*, [d]**

Department of Mechanical Engineering, Indian Institute of Technology Madras

Chennai – 600036, India

[a] E–mail: lakshmisirisha.maganti@gmail.com , [b] E–mail: pdhar1990@gmail.com

\* *Corresponding authors:* [c] E–mail: tsundar@iitm.ac.in , [d] E–mail: skdas@iitm.ac.in

## Abstract

Fluidic maldistribution in microscale multichannel devices requires deep understanding to achieve optimized flow and heat transfer characteristics. A thorough computational study has been performed to understand the concentration and thermo–hydraulic maldistribution of nanofluids in parallel microchannel systems using an Eulerian–Lagrangian twin phase model. The study reveals that nanofluids cannot be treated as homogeneous single phase fluids in such complex flow domains and effective property models fail drastically to predict the performance parameters. To comprehend the distribution of the particulate phase, a novel concentration maldistribution factor has been proposed. It has been observed that distribution of particles need not essentially follow the flow pattern, leading to higher thermal performance than expected from homogeneous models. Particle maldistribution has been conclusively shown to be due to various migration and diffusive phenomena like Stokesian drag, Brownian motion, thermophoretic drift, etc. The implications of particle distribution on the cooling performance have been illustrated and *smart fluid* effects (reduced magnitude of maximum temperature) have been observed and a mathematical model to predict the enhanced cooling performance in such flow geometries has been proposed. The article presents lucidly the effectiveness of discrete phase approach in modelling nanofluid thermo–hydraulics and sheds insight on behavior of nanofluids in complex flow domains.

# 1. Introduction

In the modern era, miniaturization of microelectronic devices and systems coupled with increased functionalities poses severe challenges to cooling technologies due to generation of high heat fluxes. Conventional cooling techniques prove inadequate in such cases and might lead to device failure due to improper thermal management. Parallel microchannel based heat exchanger devices, where a cooling fluid flows through a large number of parallel, micro machined or etched conduits, is becoming the preferred cooling device to cool modern electronic components like MEMS, VLSI circuits, laser diode arrays, high–energy mirrors and other compact products emitting high transient thermal loads. The microscale flows ensure higher levels of absorption of energy per unit volume and also provide enhanced values of convective heat transfer coefficient per unit volume and have thus been a major focus for thermo–fluidics researchers over the last two decades. In a pioneering work, Tuckerman and Peace [1] proposed a novel cooling technique using microchannel heat exchangers which are capable of dissipating large amounts of heat from small areas with high heat transfer rates and less operating fluid requirements. Later, several researchers stressed upon the applicability of conventional fluidics theories on microchannel flow domains [2–5] and it has been shown that the classical Navier–Stokes equations can be utilized for accurate prediction of liquid flow characteristics in microchannels. Though some discrepancies remain, these have been associated to factors such as measurement inaccuracies, imperfections induced during test section and geometry fabrication, entrance, exit and bend effects and effects of surface roughness. However, despite all such positives, the overall thermal performance of parallel microchannel cooling systems can be reduced because of non–uniform distribution of the working fluid from the manifold to the channels. Thereby it becomes an utmost necessity to properly understand the flow maldistribution behavior in such systems since grossly non–uniform cooling can lead to failure of certain regions of the source device. The extent of flow maldistribution in macro and mini–channels are well understood from the several proposed models [6–8] however such models fail to predict maldistribution of flow in parallel microchannels [9] since such models either neglect frictional effects within channels or the inertial effects in the manifold while both effects are equally important in case of parallel microchannels [9]. There are several experimental and numerical reports that attempt to understand flow distribution of single phase flows in parallel microchannels [10–13], for both adiabatic and heat transfer cases. Based on experiments and computations, Siva et al. [14] proposed an optimum configuration to reduce single phase flow maldistribution in parallel microchannel cooling systems.

Later, the attention shifted towards obtaining higher thermal transport by modification of the flow field or the fluid itself so as to bring in the practical implementation aspects, such as enhancement of heat transfer using offset fins or employing nanofluids as the working fluid [15, 16]. Nanofluids, which are engineered dilute and stable colloidal suspensions of metallic and/or ceramic nanoparticles in a conventional base fluid, exhibit thermal conductivity values ~ 20–150 % higher than the base fluids [17]. Several experimental and some theoretical works have been reported on the enhanced thermal conductivity of nanofluids [18–21] over the past decade. The thermal transport caliber of any nanofluid depends mainly on nanoparticle concentration, thermal conductivity, the diameter of particles, base fluid conductivity and temperature [22]. Several studies [23–25] have conclusively reported that nanofluids show great promise for use in cooling technologies. The use of nanofluids in microchannel heat exchangers has been recommended as a potentially feasible solution for cooling microelectronic devices. There are several experimental and numerical reports that concentrate on understanding the enhanced heat transfer characteristics and pressure drop of nanofluids in parallel microchannel systems [26–32]. It has been reported that enhanced heat transfer can be achieved with the use of nanofluid in microchannels but at the cost of increased pressure drop. Further, the mechanisms involved in the heat transport phenomena are not fully understood and may need more analysis [33, 34]. Overall there are few reports which concentrate on the modeling of flow and heat transfer characteristics of with nanofluids in microchannels but all these consider nanofluids as homogeneous single component fluids for analysis; which has been conclusive reported [16] to be an inefficient and incorrect assumption. Thorough survey of literature reveals there are no reports which try to understand the effects of flow and particle concentration distribution of nanofluids (treated as non–homogeneous twin component fluids) in parallel microchannels and its impact vis–à–vis thermal capabilities and uniformity. So there is a need to carry out an in–depth study to understand the effects of nanofluid maldistribution along with nanoparticle concentration and temperature maldistribution in parallel microchannel cooling systems since such a study may directly contribute towards design and optimization of nanofluid properties and microchannel systems for increasing the performance of parallel microchannel cooling systems employing nanofluids.

## 2. Numerical formulation

To understand the concentration and thermo-hydraulic maldistribution of nanofluids within parallel microchannels, detailed numerical investigation on the flow and heat transfer of alumina–water nanofluid in parallel microchannel system has been carried out. There are two different approaches used in the present work, Effective Property Modeling (EPM) and Discrete Phase Modeling (DPM) (Eulerian–Lagrangian approach). The former one considers the nanofluid as single phase homogeneous fluid with effective physical properties which are linear functions of fluid and particle material properties. The latter considers nanofluid as two phase non–homogeneous fluid i.e., fluid phase as continuous phase with nanoparticles as a discrete dispersed phase and considers all the prevalent diffusion and migration mechanisms of the nanoparticles within the fluid, viz. hydrodynamic forces, Brownian and thermophoresis diffusion, shear induced migration, etc. The present work focuses on elaborating why the DPM is a must requirement to model nanofluid behavior in microchannel systems.

### 2.1. Governing equations for the continuous phase

The governing equations for the EPM and continuous phase of the DPM are the continuity equation (mass), Navier–Stokes equation (momentum) and energy equation. The following equations respectively represent the mathematical formulations for the same.

$$\frac{\partial \rho}{\partial t} + \nabla . (\rho \vec{V}) = 0 \qquad (1)$$

$$\frac{\partial \rho \vec{V}}{\partial t} + \nabla . (\rho \vec{V} \vec{V}) = -\nabla P + \nabla . \left( \mu (\nabla \vec{V} + \nabla V^T) \right) + S_m \qquad (2)$$

$$\rho C \left[ \frac{\partial T}{\partial t} + \vec{V} . \nabla T \right] = \nabla . [k \nabla T] + S_e \qquad (3)$$

The effects of viscous dissipation and work due to compressibility are assumed to be negligible in the energy equation. In Eqns. (1)–(3), $\rho$ is density of liquid, $V$ is velocity of the liquid, $t$ is time, $P$ is pressure, $g$ is the acceleration due to gravity, $C$ is the specific heat of fluid, $k$ is thermal conductivity of fluid and $T$ is fluid temperature. $S_m$ and $S_e$ are source terms representing momentum and energy exchange respectively between the continuous phase (fluid) and discrete phase (nanoparticles) and the terms are zero for single phase model i.e. EPM.

## 2.2. Governing equations for the dispersed phase

The particle trajectories in the flow field are determined by Newton's second law of motion. Considering a Lagrangian frame of reference, the governing equation (in Cartesian coordinates) for the motion of the nanoparticles is expressed as

$$\frac{dV_p}{dt} = F \tag{4}$$

$$F = F_D + F_G + F_B + F_T + F_L + F_P + F_V \tag{5}$$

Where $V_p$ is the instantaneous velocity of the particles and $F$ is the net specific force acting on the particle. The terms $F_D$, $F_G$, $F_B$, $F_T$, $F_L$, $F_P$ and $F_V$ represent the forces due to fluidic drag, gravity, Brownian motion, thermophoretic drift, Saffman lift, contribution due to pressure gradient and contribution due to virtual mass respectively. The forces can be expressed mathematically as follows [35]

$$F_D = \frac{18\mu}{\rho_p d_p^2} \frac{C_D Re}{24} \tag{6}$$

For submicron particles as is the present case, the classical form of Stokesian drag needs to be modified so as to accommodate the non–continuum or slip boundary effects (which creeps in for high Knudsen number systems, such as flow past nanoscale particles) at the particle–fluid interface and can be expressed as

$$F_D = \frac{18\mu}{\rho_p d_p^2 C_c} \tag{7}$$

Where $C_c$ represents the Cunningham correction factor to Stokes law and the expression for the same is as

$$C_c = 1 + \frac{2\lambda}{d_p}\left(1.257 + 0.4 e^{-(1.1 d_p/2\lambda)}\right) \tag{8}$$

$$F_G = \frac{g(\rho_p - \rho)}{\rho_p} \tag{9}$$

Since Brownian motion is random in nature with zero net directional flux, a probability function is required to model the force. The amplitude of the Brownian force components is expressed as

$$F_{B_i} = \zeta_i \sqrt{\frac{\pi S_0}{\Delta t}} \tag{10}$$

where $\zeta_i$ is a random number which is part of a Gaussian distribution with zero mean. The amplitudes of the Brownian force components are estimated at each step of the discrete phase calculations. The components of the Brownian randomness are modeled as Gaussian white noise process with the expression for the spectral intensity $S_{n,ij}$ expressible as [36]

$$S_{n,ij} = S_0 \delta_{ij} \tag{11}$$

where $\delta_{ij}$ is the Kronecker delta function and the expression for the amplitude of the spectrum $S_0$ is expressed as

$$S_0 = \frac{216 \nu k_B T}{\pi^2 \rho d_p^5 \left(\frac{\rho_p}{\rho}\right)^2 C_c} \tag{12}$$

The dispersed particles within a continuous phase subjected to a temperature gradient experience a force in the direction opposite to that of the gradient due to higher degree of molecular bombardment on the particles at the heated region, driving it towards the colder region where the net force due to bombardment is less. The phenomenon is known as thermophoresis or Soret effect and the expression for the force generated due to the drift is expressed as

$$F_T = -D_{T,P} \frac{1}{m_p T} \frac{\partial T}{\partial x} \tag{13}$$

Where $D_{T,P}$ is the thermophoretic coefficient [37]

$$D_{T,P} = \frac{6\pi d_p \mu^2 C_s (k + C_t k_n)}{\rho (1 + 3 C_m k_n)(1 + 2k + 2 C_t k_n)} \tag{14}$$

Where $C_m=1.146$, $C_s=1.147$ and $C_t=2.18$ are the momentum exchange, thermal slip and temperature jump coefficients respectively.

The Saffman lift force which is generated due to shear on the particle by the continuous phase (this form of lift arises for small particles in flow) is expressed as

$$F_L = \frac{2 k_s \nu^{\frac{1}{2}} \rho d_{ij}}{\rho_p d_p (d_{lk} d_{kl})^{1/4}} (V - V_p) \tag{15}$$

where $k_s = 2.594$ is a constant and $d_{ij}$ is the deformation tensor for the continuous phase which governs the shear generated around the particle.

The force arising on the particles due to pressure gradient within the fluid is expressed as

$$F_P = \left(\frac{\rho}{\rho_p}\right) V_p \frac{\partial V}{\partial x} \tag{16}$$

The inertia required to propel the fluid surrounding the particles gives rise to a virtual mass force and can be expressed as

$$F_V = \frac{1}{2} \frac{\rho}{\rho_p} \frac{d}{dt}(V - V_p) \tag{17}$$

## 2.3. Effective Property Model

The following formulations have been used for determining the effective properties (density, specific heat, viscosity and thermal conductivity in ascending order of equation numbers) of alumina–water nanofluid considering such fluids as homogeneous single component systems [38]

$$\rho_{nf} = (1 - \phi)\rho_{bf} + \phi\rho_p \tag{18}$$

$$(\rho C_p)_{nf} = (1 - \phi)(\rho C_p)_{bf} + \phi(\rho C_p)_p \tag{19}$$

$$\mu_{nf} = \mu_{bf}(1 + 10\phi) \tag{20}$$

$$k_{eff} = k_f \frac{[k_p + (n-1)k_{BF} - (n-1)\phi(k_{BF} - k_p)]}{[k_p + (n-1)k_{BF} + \phi(k_{BF} - k_p)]} \tag{21}$$

## 2.4. Computational details

A 3–D, U type, parallel microchannel domain has been created, meshed and fluid flow and heat transfer solved employing ANSYS Fluent 14.5. Fig. 1(a) shows the geometrical configuration utilized in the present study. This particular geometry has been revealed to have the worst flow distribution characteristics [9] and hence studies on the same provide information on nanofluid flow in microchannels for the worst case scenario; an essentiality for design and optimization. The details of dimensions of geometry and working fluid are as follows: hydraulic diameter ($D_h$) of channel is 100μm, area ratio ($A_{channel}/A_{manifold}$) is 0.2, number of channels (N) is 7, aspect ratio of channel (H/W) is 0.1, working fluid is water and $Al_2O_3$–water nanofluid. A mesh consisting of quadrilateral elements has been utilized and employs the grid at the inlet of the manifold for injecting the nanoparticles. A grid independence study is carried out by considering different mesh element numbers and Fig. 1 (b) shows the grid independent study results considering the flow maldistribution parameter (expressed in Eqn. 22) criteria [9] for grid independence test. As evident from the figure, there is no change in maldistribution parameter with respect to number of mesh elements beyond 1250000. A finer element size (1455237 number of mesh elements) is considered for the present study since availability of large number of surfaces at inlet to inject more particle streams renders tracking more accurate. Uniform heat flux has been applied at the bottom and side walls for heat transfer cases and the top wall has been considered adiabatic.

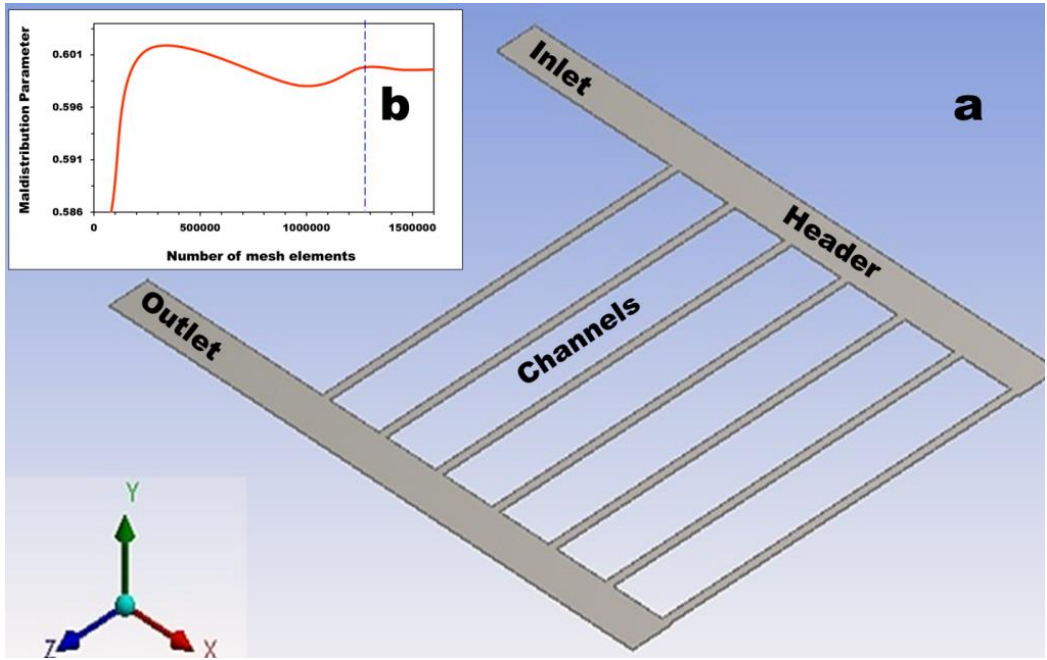

**Figure 1:** **(a)** Geometry of parallel microchannel system used as the simulation domain **(b)** Grid independence test (maldistribution parameter with respect to number of mesh elements).

The present numerical model has been validated with respect to the published reports by Siva et al. [9] and Singh et al. [16]. The former study discusses in details flow maldistribution of water in parallel microchannel systems whereas the later study comprises detailed report on the thermo–hydraulic performance of nanofluids in single microchannel system. The present study being an effort to shed insight on to the flow and thermal behavior of nanofluids in parallel microchannel systems is thereby justifiably validated from the two mentioned sources and the plots have been illustrated in Fig. **2**(a), (b) and (c). Fig. 2 (a) validates the present microchannel model against published data [9], wherein the maldistribution of water among the parallel channels for two different hydraulic diameters (88μm and 176μm and flow at Re=70) has been considered. It can be observed from Fig. 2 (a) that the present simulations accurately track the reported results and this paves a roadway for justifying the homogeneous model results for nanofluids in parallel microchannels (presented in later sections). Fig. 2 (b) and (c) illustrates the efficacy of the present model in simulating flow and thermal transport compared to the experimental reports [16]. The experiments report flow and heat transfer in nanofluids within a single microchannel and illustrates the effectiveness of the Eulerian–Lagrangian particle tracking models in such twin phase flows. It is evident from the figure that the present Discrete Phase Model (DPM) agrees well with reported experimental investigations. Consequently, validating against documented experimental data for flow of simple fluids and nanofluids in both single as well as multiple microchannel assemblies essentially provides evidence that the present model can effectively simulate both homogeneous and discrete phase approaches to determine performance of nanofluids in microchannel cooling and to provide insight onto the associated flow and thermal physics.

## 3. Results and Discussions:

### 3.1. Adiabatic flows

#### 3.1.1. Pressure drop and flow maldistribution:

In order to comprehensively project the performance of nanofluids as potential coolants in microelectronics or micromechanical devices employing parallel microchannel systems, it is of utmost importance to first shed light onto the adiabatic transport of the same. While in case of simple and/or single phase fluids the major adversity to be addressed or modified is the hydraulic maldistribution in the channel systems, in case of complex and non–homogeneous fluids such as nanofluids, maldistribution of the effective concentration is also expected to pose additional concerns towards performance of such systems. Thereby it deems a necessity that a detailed Eulerian–Lagrangian particle tracking model be employed to simulate such flows and establish the deviances from the homogeneous property models. Furthermore, it is pertinent that the flow regimes be identified for the system geometry under consideration within which such maldistribution is appreciably high and sensitive to changes in flow Reynolds number. Accordingly, the effects of Reynolds number and concentration on flow and concentration maldistribution of nanofluids in parallel microchannel systems have been numerically investigated using the DPM. The flow maldistribution has been quantified based on the flow maldistribution factor (FMF) expressible as [9]

$$\eta = \left(1 - \frac{\Delta P_{min}}{\Delta P_{max}}\right) \qquad (22)$$

Similarly, the extent of concentration maldistribution is quantified using the concentration maldistribution factor (CMF), defined as

$$\varepsilon = \left(1 - \frac{\phi_{min}}{\phi_{max}}\right) \qquad (23)$$

The magnitudes of the FMF and CMF vary between 0 and 1, where 1 represents a scenario of maximal maldistribution.

The present study utilizes generalized nanofluid formulation throughout and owing to excellent transport characteristics and stability, aluminum oxide (40–50 nm) and water based nanofluids have been used [23]. Furthermore, a basic U–type manifold and channel geometry is considered as it has been reported to exhibit highest maldistribution (compared to I and Z configurations, [9] and hence a clear picture of nanofluid performance in the worst case scenario can be obtained. Channel wise pressure drop, a parameter important to characterize flow features and pumping requirements in parallel channel systems, has been illustrated in Fig. 3 (a), for nanofluid at three different concentrations (1, 3 and 5 vol. %) and for two different Reynolds numbers (2 and 50; one low another moderately high). As evident from the figure, the pressure drop across the channels is higher for the nanofluid compared to water, which is expected given the higher viscosity of the nanofluid induced by the presence of nanostructures within the fluid. The pressure drop in the initial channels is higher for both water and nanofluid when compared to those in the later channels due to the non–uniform distribution of fluid in the parallel microchannels; the effect known as flow maldistribution. However, knowledge of the pressure drop values in the channels individually does not portray a complete picture onto the maldistribution characteristics within the overall geometry. The extent of maldistribution for water and nanofluids has been illustrated in Fig. 3(b) by the maldistribution parameter (η) at different Reynolds numbers and for different concentrations.

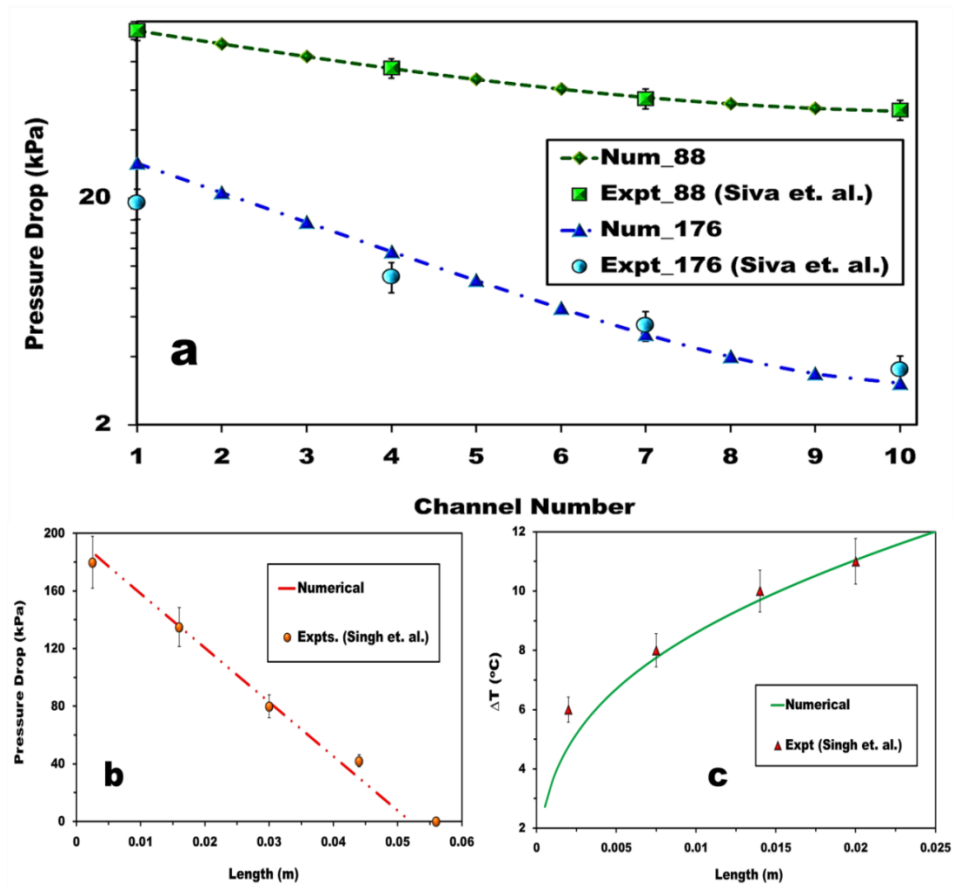

**Figure 2:** Validation of present numerical model with reported experimental investigations **(a)** Validation with Siva et al. experimental results **(b)** Validation with Singh et al. experimental results for adiabatic transport **(c)** Validation with Singh et al. experimental results for diabatic transport.

It can be observed from the figure that hydraulic maldistribution increases gradually as a function of nanofluid concentration and the effect is further enhanced at lower Reynolds numbers. However, in reality, enhanced viscosity is expected to induce more uniform distribution and the enhanced flow maldistribution at higher concentration thus provides the first hint at the behavior of nanofluids as complex, non–homogeneous fluids, where the distribution of the particles governs the flow behavior. At high Reynolds numbers, the flow is dominated by inertia, enabling the later channels more share of the working fluid which in turn reduces maldistribution. Due to high inertial effects, the shear and diffusion induced migration of the nanoparticles is arrested and the particles are forced to track the streamlines along the direction of flow, and accordingly, the FMF becomes independent of nanofluid concentrations at high flow velocities. However, at low Reynolds numbers, the inertia of flow is less and hence resistance to the random motion of particles due to Brownian effect and shear induced migration is less. Thereby, the enhanced motion of the nanoparticles leads to concentration maldistribution, which in turn affects the localized viscous forces and causes further maldistribution of flow. It can be thus inferred that flow maldistribution exhibits sensitivity to particle concentration and the deviation from the base flow increases with increasing particle concentrations at low Reynolds numbers (flow regimes expected in real scenario applications of microscale flow based heat transfer devices). Nanofluids will therefore not behave as homogeneous fluids in such devices and hence their transport capabilities in microchannel systems cannot be predicted by conventional numerical methods employing Effective Property Models (EPM) wherein the nanofluid is treated as a homogeneous, single component fluid.

Further insight into the behavior of nanofluids in such complex flow paths can be assessed from comparison of maldistribution obtained from DPM and EPM analyses, as illustrated in Fig. 4. As observable, the FMF predicted by the EPM remains independent to changes in either concentration or Re, except for highly concentrated fluids and this anomaly arises due to the EPM's treatment of nanofluids as homogeneous and single component, wherein fluid properties are calculated based on effective material properties. From Fig. 4 it can be observed that the EPM FMF at 1 and 3 vol. % are similar in magnitude and this occurs due to the usage of expressions such as Einstein's or Batchelor's' equations [39] for determining viscosity of suspensions in the EPM. These expressions work well only for very dilute suspensions and the predictions are weakly dependent on concentration, which leads to similar viscosity values in the two cases, leading to similar FMF. However, at 5 %, the viscosity value predicted by the EPM increases marginally, leading to marginal drop in the FMF, but all the predictions remain independent of Re since the distribution of the single phase nanofluid is unaffected by the inertial effects in the range considered. On the contrary, the variation of FMF can be observed clearly as functions of Re and concentration when DPM is resorted to and the observations are credible as Eulerian–Lagrangian approach of modeling nanofluids has been reported to predict experimental observations unlike its single phase counterparts.

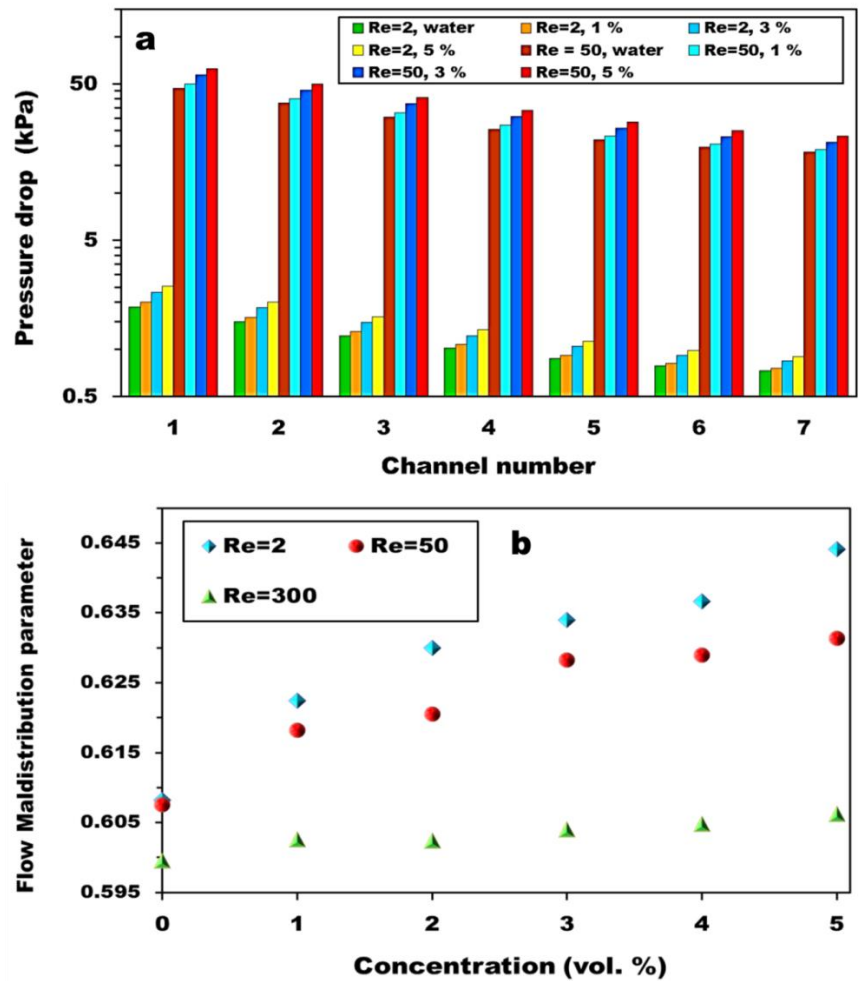

**Figure 3:** **(a)** Comparison of nanofluid pressure drop across each channel for three different concentrations (1, 3 and 5 vol. %) with water **(b)** Behavior of FMF (η) with respect to concentration at three Re (in the low, moderate and high inertial regimes).

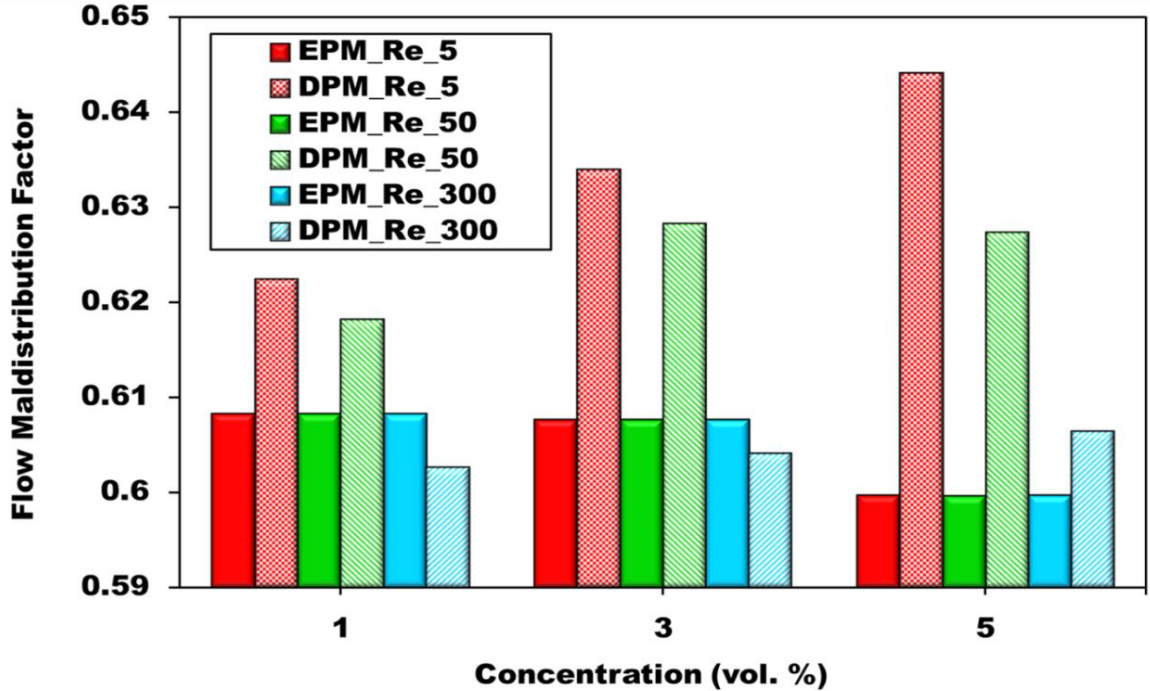

**Figure 4:** Comparison of FMF of nanofluid obtained utilizing DPM and EPM approaches at three different concentrations and for three different Re.

The DPM approach is able to capture the proper FMF as the model considers the particles as a phase in some ways independent of the fluid phase and tracks the migration of the particles (considering all the diffusive effects like Brownian fluctuations, Saffmann lift, thermophoresis, Stokesian drag, rotation and so on) within the continuous phase and its interactions with the fluid as well as neighboring particles. It can be observed in Fig. 4 that increment in concentration at a particular Re leads to increased DPM FMF, as opposed to the decreasing trend in EPM. Enhanced particle population expectedly enhanced the viscosity of the nanofluid, which in accordance to EPM should lead to reduced FMF. However, the fact that increased particle count per unit volume introduces higher degree of Brownian fluctuations and more importantly drag, are taken into consideration by the DPM. Exemplary scenario for the enhanced maldistribution can be provided at this instance. If the first channel be considered, the fluid component of the nanofluid gets distributed similarly to that of the base fluid. However, owing to higher inertia of the particles (due to the higher density), only a small fraction of the particle enter the first channel and effectively enhance the concentration of the fluid heading to the next channel. This enhanced and varying load thereby prevents the fluidic phase to be distributed similar to the base fluid in the later channels and this continues so forth, thereby inducing higher degrees of maldistribution to the flow. With increasing concentration, this effect enhances, leading to further hampering of flow distribution. Increase in flow Re leads to deceased FMF and this is caused by the dominance of flow inertia. At higher flow velocities, the diffusive and migration effects of the particles decrease and they more or less follow the flow pattern, leading to more uniform distribution. In fact, the DPM FMF approaches the EPM FMF as Re increases, providing evidence that the nanofluid behavior asymptotically approaches homogeneous fluid behavior at high inertia regimes.

### 3.1.2. Concentration maldistribution

As discussed in the preceding section, it is also important to understand the particle concentration distribution during nanofluid flow in parallel microchannels and is something that all published reports in the field have overlooked. Since it directly affects the cooling performance, a comprehensive understanding can provide better suited design approaches for nanofluid based microchannel heat exchangers. Common intuition, considering nanofluids similar to single phase systems suggests that the nanofluid should distribute similar to the base fluid, however, this is far from the reality. Fig. 5 illustrates a comparison between the FMF and concentration maldistribution factor (CMF) for different concentrations and Re. As discussed, it can be inferred from the figure that nanofluids do not behave like homogeneous fluids as the FMF and CMFs are grossly dissimilar at different Re and concentrations. While the trends of both flow and concentration maldistributions as function of inlet concentration are qualitatively similar at low Re, they are absolutely different at high Re. In fact, Fig 5 provides further evidence as to the failure of the EPM and the process by which the maldistribution of concentration in turn leads to non–intuitive flow maldistribution can be gauged. While the FMF is expected to reduce for concentrated nanofluids, the reverse occurs.

At low Re, the particles are more independent to migrate and diffuse across the streamlines, and this in turn leads to non–uniform distribution of concentration. As the particle loading increases, the migration effects, fluidic drag and inter–particle interactions increase, leading to higher CMF. This in turn affects the flow and the FMF enhances too, as discussed in the preceding section. As Re increases, EPM predicts no noticeable changes in the FMF than that of low Re, however, DPM predicts appreciable changes. While the decrease in FMF compared to low Re scenario can be justified based on the higher inertia of flow which arrests particle migration to some extent; the decrease of CMF at higher Re with increasing concentrations needs deeper insight. With increasing Re for the same hydraulic diameter, the flux of the fluid increases and accordingly the streamlines are packed closer. In such cases, although inertia has arrested diffusive movements orthogonal to the streamlines severely, the particles still have scope to diffuse and migrate along the direction of the flow. This effect still leads to uneven distribution and hence at low particle populations, the CMF remains fairly unaffected. However, as the concentration is increased, the population is packed within the closely placed streamlines and the migratory movements along the streamlines are also cut off due to excessive particles in the system. The system thus begins to behave like a packed bed of granular media and flows more or less along with the base fluid, thereby reducing the concentration maldistribution. This effect is further pronounced at higher Re values and the CMF at high concentration further decreases.

The distribution of the particles within the flow geometry also requires a qualitative analysis so as to understand the overall behavior of nanofluids in microchannel systems. The DPM concentration profiles at the horizontal geometry mid–plane at low Re for three different concentrations have been illustrated in the Fig. 6 (a), (b) and (c). As discussed earlier, at low Re, the particle concentration distribution is relatively uniform at low concentrations compared with high concentrations and this can be seen qualitatively from Fig. 6. At low concentrations, it can be observed that a large fraction of the population is channelized through channel 1, followed by channel 2, whereas the later channels experience flows of much reduced concentration. With increased concentration to 3 %, the scenario improves with the $3^{rd}$ and $4^{th}$ channels getting a fair share of particles. This happens expectedly as a major fraction of the increased population cannot travel through the $1^{st}$ and $2^{nd}$ channels completely. As the concentration is further increased, the end channels also start experiencing a large fraction of the particles. In fact, as discussed earlier, movement analogous to that of a packed bed leads to higher concentration flows within the central and end channels.

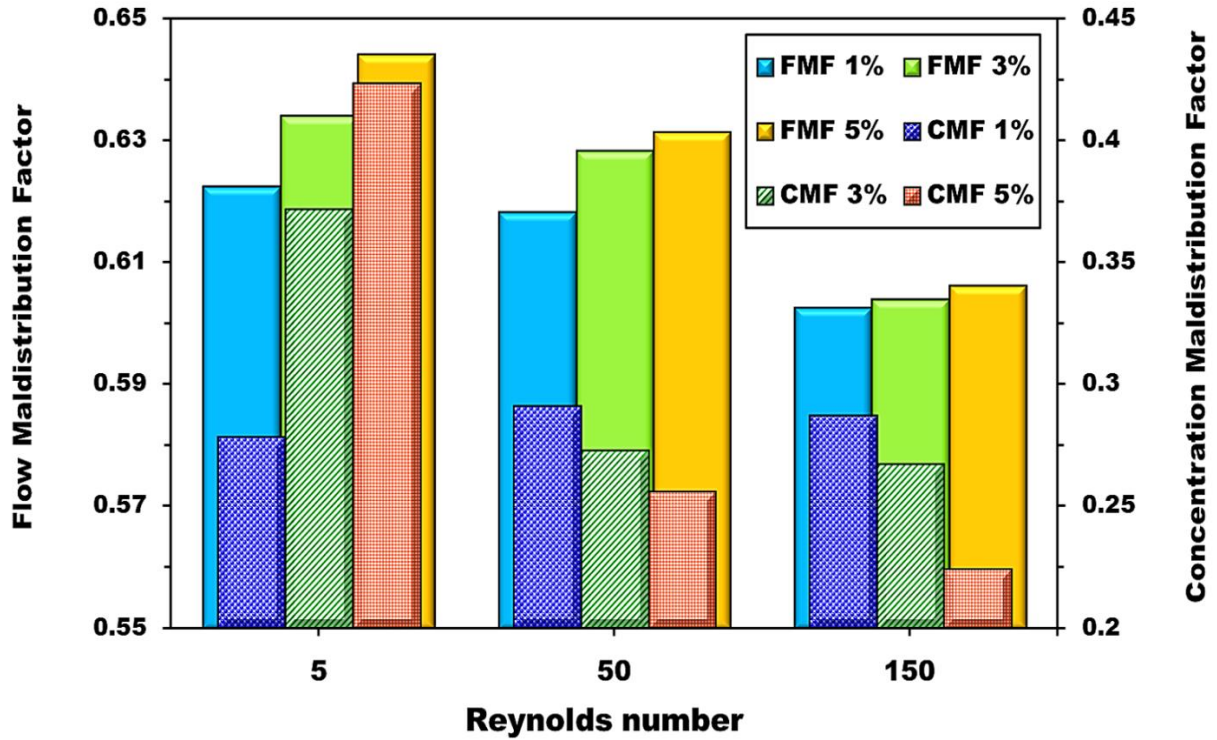

**Figure 5:** Comparison of DPM FMF and CMF for the nanofluid at three different Re and concentrations.

The concentration distribution contours at a cross section of the inlet manifold (as shown by arrow) and a cross section of the outlet manifold (as shown by the arrow) have been illustrated in Fig. 6(b1) and (b2). It can be observed that while the area weighted mean concentration of the two sections remain same (preservation of continuity of the discrete phase); the distribution patterns are grossly different. While the distribution at the inlet manifold consists of many regions of concentrated zones of particle population, its outlet counterpart consists of a more diffused concentration distribution. In the inlet region, the sole flow mechanism that actuates mixing of the particulate phase is the convergence of the boundary layers within the developing region. Within the developing region the mixing is null in the potential flow zone and full scale mixing begins only convergence and establishment of complete viscid flow regime. However, the outlet manifold contains flow already experienced to the effects of entrance, exit and bend of the flow and to mixing of different merging streams of different effective concentrations. Accordingly, the discrete phase is much more diffused and well dispersed within the outlet manifold than the inlet.

An accurate qualitative assessment of the impact of the particle slip forces on the concentration maldistribution can be made from the maldistribution pattern at sections very near (within a few grid lengths) the entrance of the inlet manifold. The concentration distribution contours at a section proximal to the inlet cross section at different Re have been shown in Fig. 6(d). A non–uniform concentration distribution can be observed to prevail at the entrance of the inlet manifold at low Re and the uniformity of concentration distribution improves as Re increases. The diffusion or migration of particles away from the point of entry at regions very near the entrance of the manifold is due to Brownian motion, since in this region it is the only slip mechanism which is existent (at the inlet flow is yet to be established and hence drag, lift etc. are not present). At very low Re, the inertia of the continuous phase is small in magnitude and the Brownian velocity of particles is comparable with the continuous phase velocity. Hence, diffusion or migration of the particles away from the streamlines takes place spontaneously; leading to non–uniform distribution of concentration at the entrance of the manifold itself. This effect perishes as the Re increases and the phenomenon is observed only when ratio of continuous phase velocity to Brownian velocity is below 500 ($V_C / V_B <$ 500). To justify the above observations, simulations have also been carried out by switching–off the Brownian component in the governing equations and the corresponding results have been illustrated in Fig. 7. Fig. 7(a) and (b) exhibit the concentration distribution contours at entrance and exit of the inlet and outlet manifolds respectively without the Brownian effect. Fig.7(c) and (d) illustrate the same with the Brownian effect incorporated. As observable, while the outlets show some similarities in the distribution pattern, the inlets are grossly dissimilar and the effect of Brownian motion on particle maldistribution can be comprehensively understood, thereby making it one of the most important phenomena at low Re flows of nanofluids in microscale flow devices.

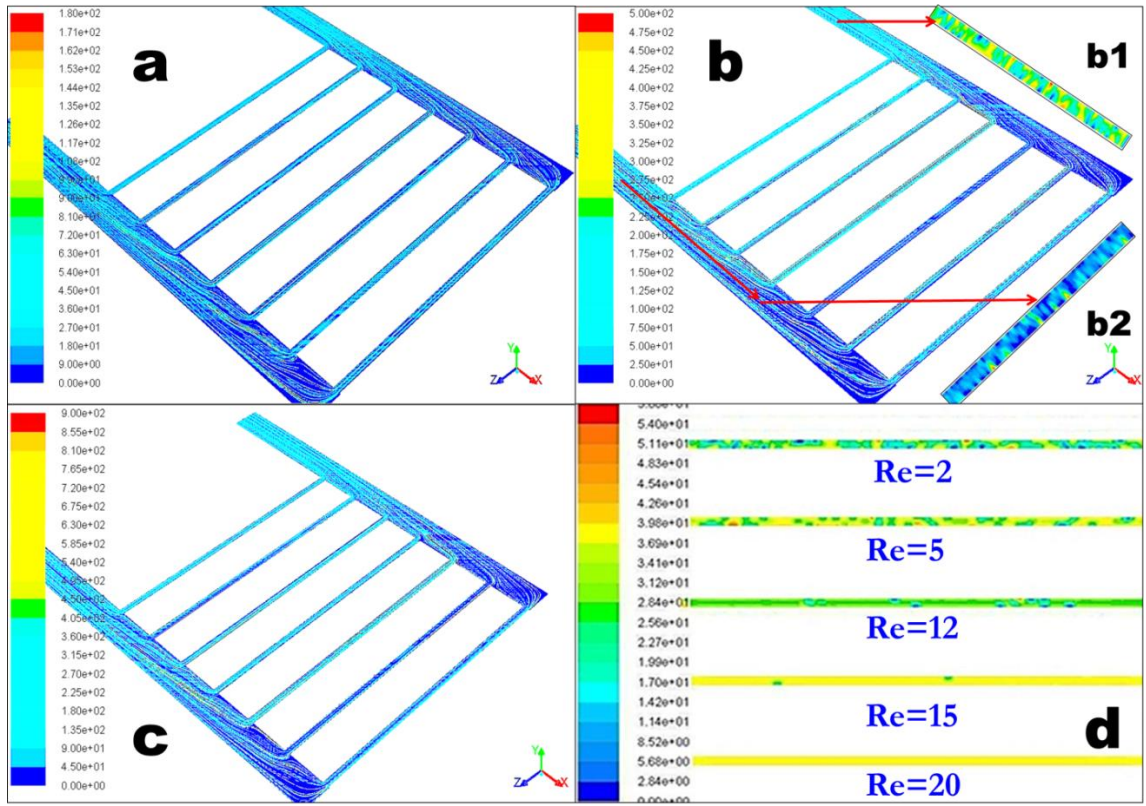

**Figure 6:** Contours of dispersed phase concentration of the nanofluid among the parallel microchannels at Re=2 for different concentrations **(a)** ϕ = 1 vol. % **(b)** ϕ = 3 vol. % **(b1)** Concentration distributions at inlet manifold cross section **(b2)** Concentration distributions at outlet manifold cross section **(c)** ϕ = 5 vol. % **(d)** Contours of concentration distribution at inlet cross section of inlet manifold at different Reynolds numbers for 1 vol. % with Brownian diffusion active within the DPM formulation.

The effect of flow inertia on the distribution of the nanofluid and its implications vis–à–vis concentration maldistribution among the individual can be assessed from the concentration contours within specific channel inlets for different Re. Fig. 8 illustrates the cross sectional concentration contours at regions very near the inlets of channels 3, 5 and 7 for three different Re. At low Re, the inertia of the fluid within the inlet manifold is low, thereby allowing the front channels to get a fair share of the particle population than the case at higher Re, where majority of the population is flushed to the later channels. This can be observed in Fig. 8, where the concentration contour in channel 3 at higher Re is much more diffused and has no particle flow aggregations as those in low Re. Channel 5, being almost within the central region, experiences very little change in distribution pattern with changing Re value. At low Re, a large extent of the particles travel into the front channels and at high Re they travel through the latter channels, leaving the central channels with fairly constant share of particles. At low Re, the last channel gets a dilute flow, as observed in the figure, where large fractions of disappearing dilution can be observed. As the Re increases, the flushing event pushes more particles to the latter channels and as evident from the figure, the distribution in channel 7 at moderate and high Re improves drastically compared to the low inertia regime. Thereby, when used in cooling technologies, probability of occurrence of hot spots can be deduced to be low among the regions housing the central channels for almost all inertial regimes.

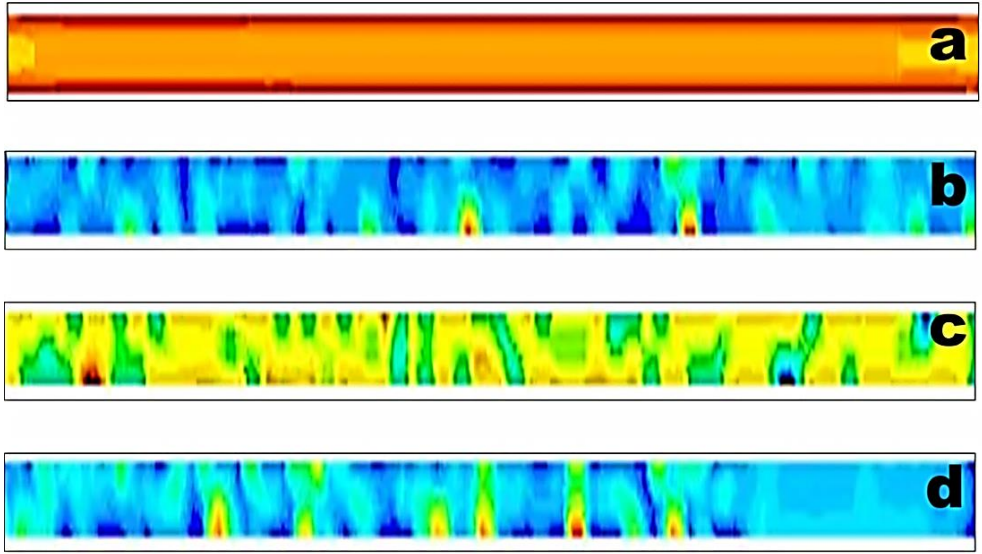

**Figure 7:** Dispersed phase mass concentration distribution contours at Re =2 at the **(a)** entrance of the inlet manifold with Brownian effect switched off **(b)** exit of the outlet manifold with Brownian effect switched off **(c)** entrance of the inlet manifold with Brownian effect incorporated **(d)** exit of the outlet manifold with Brownian effect incorporated.

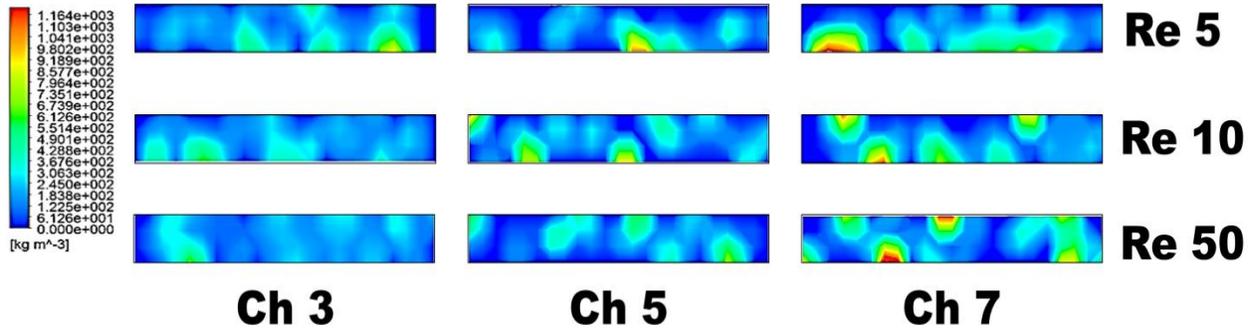

**Figure 8:** Effect of flow inertia on the concentration distribution within individual channels (at a section proximal to the channel inlet) for nanofluid of a fixed concentration.

## 3.2. Diabatic flows

### 3.2.1. Flow maldistribution

Although understanding flow maldistribution is important for optimizing the pumping characteristics, understanding the same with increasing heat loads is required for efficient design of such specialized microscale flow systems. Fig. 9 illustrates the FMF for nanofluids as function of concentration, Re and imposed heat flux. It can be observed that for the geometry considered, the presence of nanoparticles in base fluid changes the trend of fluid distribution among the parallel microchannels and the effect is more pronounced at low Re. Furthermore, the deterioration of FMF at a particular Re with increasing temperatures is more in case of the nanofluid than that of water, which brings to the forefront the important role that nanoparticle migration and diffusion (which is more prominent at elevated temperatures) in determining the overall flow pattern. As heat flux increases, the temperature in the system increases, and the viscosity of the fluid decreases, leading to increased non–uniform distribution of fluid due to enhanced inertia. However, the increment of FMF for water with respect to Re and heat flux is negligibly small. On the contrary, the FMF increases appreciably for the nanofluids (DPM simulation) with Re, heat flux and concentration and increase in FMF is more at low Re with respect to both heat flux and concentration. At low Re, as discussed earlier, resistance to the random motion of particles due to Brownian fluctuations is less and the Brownian velocity of the particles is comparable to the continuous phase velocity, leading to localized disruption of the flow field by the particle diffusion and maldistribution of fluid due to summation of the effects. As heat flux increases, the viscosity of the fluid decreases and simultaneously the thermal migration of the nanoparticles increases and the net effect lead to higher degrees of maldistribution. At high Re, inertia dominates within the flow regime and resistance to the random motion of particle is high and thus presence of particles in base fluid does not affect the distribution of fluid among the channels to appreciable extents. However, it can be observed that the FMF tends to a plateau value as the concentration increases. At concentrations beyond 5 vol. % (already in the concentrated regime), the effect of particle migration is greatly reduced by overcrowding and the viscosity of the overall fluid enhances drastically, leading to attainment of a saturation value for FMF. Several associated phenomena have been discussed in the subsequent sections where concentration maldistribution at enhanced temperatures has been dealt with in depth.

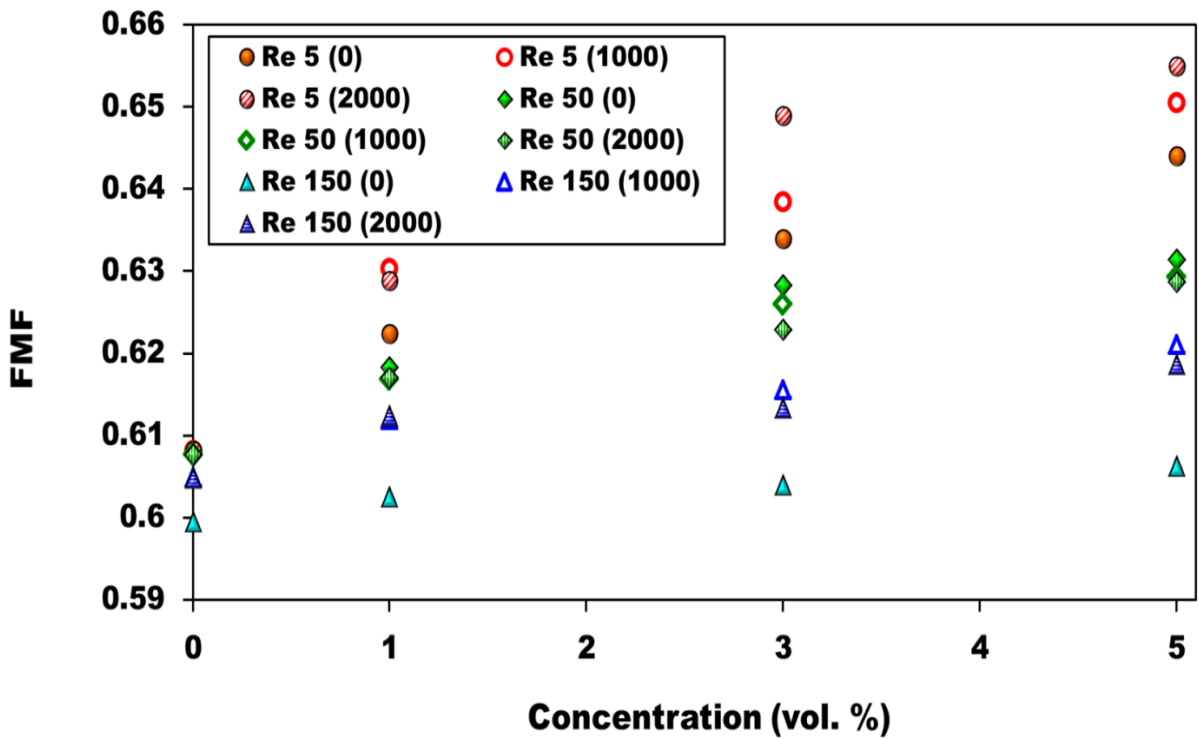

**Figure 9:** FMF for nanofluids with respect to concentration at three different heat fluxes for three different Re.

### 3.2.2. Concentration maldistribution

While understanding the flow features of nanofluids is an important aspect for the design of such microscale flow systems, a concrete understanding of the heat transport efficacy of nanofluids within such systems and the associated phenomena is of great importance. The extent of concentration maldistribution of the nanofluid as a function of Re and inlet concentration for both adiabatic and diabatic cases has been illustrated in Fig. 10 and it can be inferred that although the trends of change in CMF for adiabatic and diabatic cases remain fairly similar, they differ quantitatively. Also, several different phenomena in the distribution of particles crop up in presence of elevated temperatures. At low Re, the CMF decreases at moderate heat flux and increases marginally at high heat fluxes, with the 5 % nanofluid being the exception wherein the CMF further falls at high heat flux. At low inertia flows, upon increment of flow temperature, the decrease in fluid viscosity aids uniform distribution of the particles. While reduced viscous effects deteriorate flow distribution, the particles experience reduced drag and are thereby more free to overcome the established base flow and hence can distribute more uniformly. However, further increment in heat flux leads to further lowering of viscous drag and enhances the Brownian and thermophoretic diffusion/ migration, thereby introducing higher thermal fluctuations and disrupting the decrement in CMF to some extent.

Another way of looking at is from the particle migration point of view. In adiabatic case the random motion of the particles is solely due to Brownian diffusion whereas in diabatic case it is due to both Brownian and thermophoretic migration. Due to thermophoresis, the nanoparticles are directed away from heated channel walls and the phenomenon is more predominant at end channels because of higher temperatures due to flow maldistribution. The predominant thermophoresis may oppose the Brownian randomness, providing the particles a net directional drift that overshadows the Brownian effect. Hence, added resistance to the random motion of the particles leads to relatively more uniform distribution of concentration among the channels. At very high Re, the inertia of the flow enhances drastically with decreasing viscous forces and the flushing effect on the particles essentially enhances. However, in case of moderate Re, the CMF shoots up even for moderate heat fluxes and then reduces when the flux increases. This is in all probability caused due to sudden shift of flow regimes at such Re values. At low Re, even drastic changes in viscosity cannot be expected to transit the flow regime from predominantly viscous to high inertia. Similarly, high Re flows being predominantly inertial remain inertial due to decrease in viscosity. At moderate Re, where the flows are not dominated totally by either viscous or inertial regimes, a slight alteration in viscosity value can shift the flow to fully inertial regime, leading to drastic localized particle maldistribution. This is possibly why the CMF increases suddenly even at moderate heat fluxes. However, at high heat flux, the viscous forces further decrease and flushing in behavior slowly sets in, which reduces the maldistribution as in high Re scenarios. Also, at high Re, Brownian fluctuations are effectively arrested by inertia and thermophoretic drift reduces due to more uniform cooling at high flow velocities, resulting in similar concentration distribution among the channels for different concentrations.

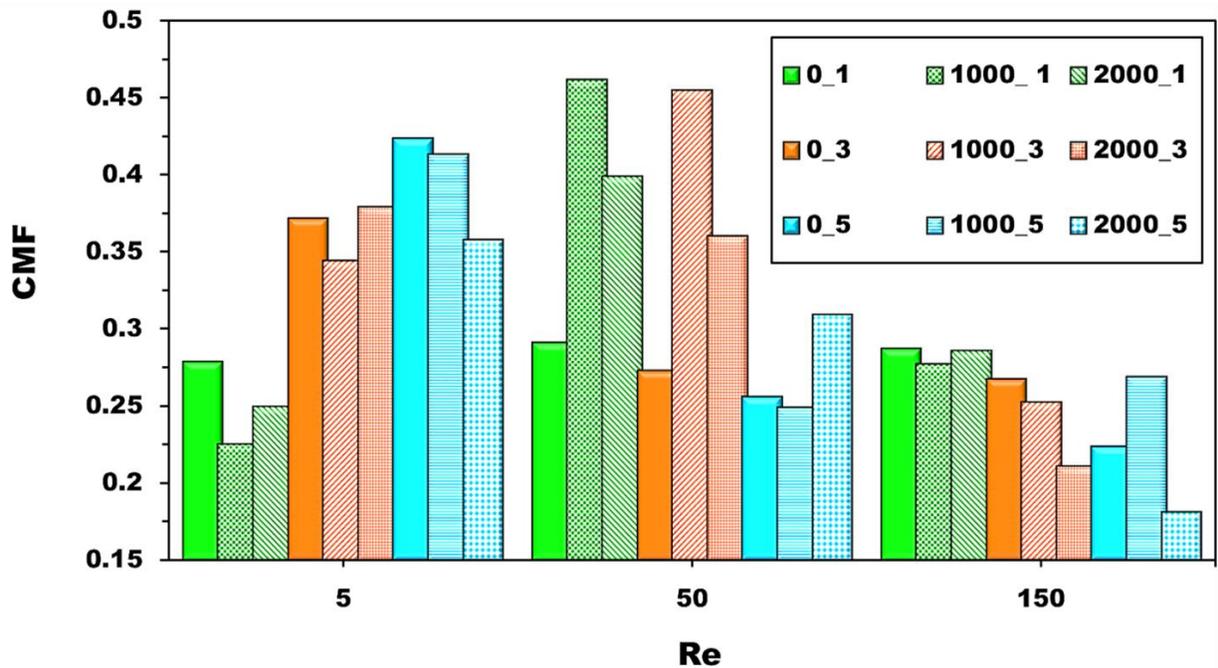

**Figure 10:** CMF for nanofluid as function of inlet concentration and Re for adiabatic flow and diabatic flows with 1 and 2 kW/m$^2$ heat fluxes applied to the geometry (legend : *heat flux _ concentration*).

A more thorough picture of the effect of temperature on the distribution of the particulate phase can be envisioned by illustrating the effective concentration of the nanofluid entering each channel under different conditions. Fig. 11 illustrates the effective concentration entering each microchannel at different heat flux levels for a base flow equivalent to Re = 5 and an effective concentration of 5 vol. % entering the inlet manifold. It can be observed from Fig. 11 that the effective concentration in the individual microchannels is different for different heat fluxes, thereby leading to differences in the CMF. Increment in temperature due to the moderate heat flux can be observed to lead to a shift in the distribution pattern. While for the adiabatic condition the former channels and the very last channel received flows of appreciable concentration, the 1 kW/m$^2$ condition leads to much better distribution among the central channels as well. As heat flux is applied, thermophoresis comes into the picture along with Brownian diffusion and since the Re is low, the resistance to the migration of the particles due to both the effects is less. Thermophoresis directs the particle population away from the manifold outer walls (due to less cooling than the channel side) and essentially towards the channels, causing the particles to distribute more uniformly among the channels than adiabatic conditions. This is in agreement to the observations if Fig. 11. For increased heat flux i.e. 2 kW/m$^2$, the location of the valleys (low effective concentration) and peaks are qualitatively similar to the adiabatic case, except for the end channels. With increment in temperature, the effect of Brownian motion increases drastically and the directionality of the thermophoretic drift is overshadowed (but not obliterated) to some extent, leading to deteriorated distribution similar to adiabatic conditions. However, towards the end of the manifold, where the temperature gradients are higher due to flow maldistribution caused by reduced viscosity, the thermophoretic drift regains upper hand and leads to better distribution than the adiabatic case.

The effect of temperature on the distribution of the particulate phase can also be qualitatively understood from the concentration contours within each channel. Fig. 12 illustrates the same at a section located at the lengthwise center of the channels for different heat fluxes and for a nanofluid of 5 vol. % and manifold flow corresponding to Re = 5. From the contours it can be observed that at 0, 1 and 2 kW/m$^2$ the maximum effective concentration exists in channel 7, channel 4 and channel 6 respectively and the minimum effective concentration can be observed in channel 5, channel 7 and channel 2 respectively (which are in agreement with the quantified data in Fig. 11a). It can further be seen that in the diabatic cases, especially for the later channels, there exist distinguishable regions of very low concentration near the side and bottom walls (such as in channels 5, 6 and 7 of 1 kW/m$^2$ and channels 6 and 7 of 2 kW/m$^2$). Such migration away from the heated channel walls is clear evidence of thermophoretic drift and is strong in the later channels as these experience large thermal gradients caused by maldistribution in the base flow. The concentration within the front channels are more diffused due to the greater degree of mixing by the base flow. As the flow moves towards the later channels, it loses inertia and the particulate phase sluggishly drifts along, forming occasional clustered regions due to lack of inertia induced mixing. However, as the viscous resistance reduces with temperature, the diffused contour can be seen to extend up to channel 4 in 2 kW/m$^2$ case as compared to channel 2 in the adiabatic case. Such observations provide firm support on the efficacy of nanofluids as future generation micro device coolants.

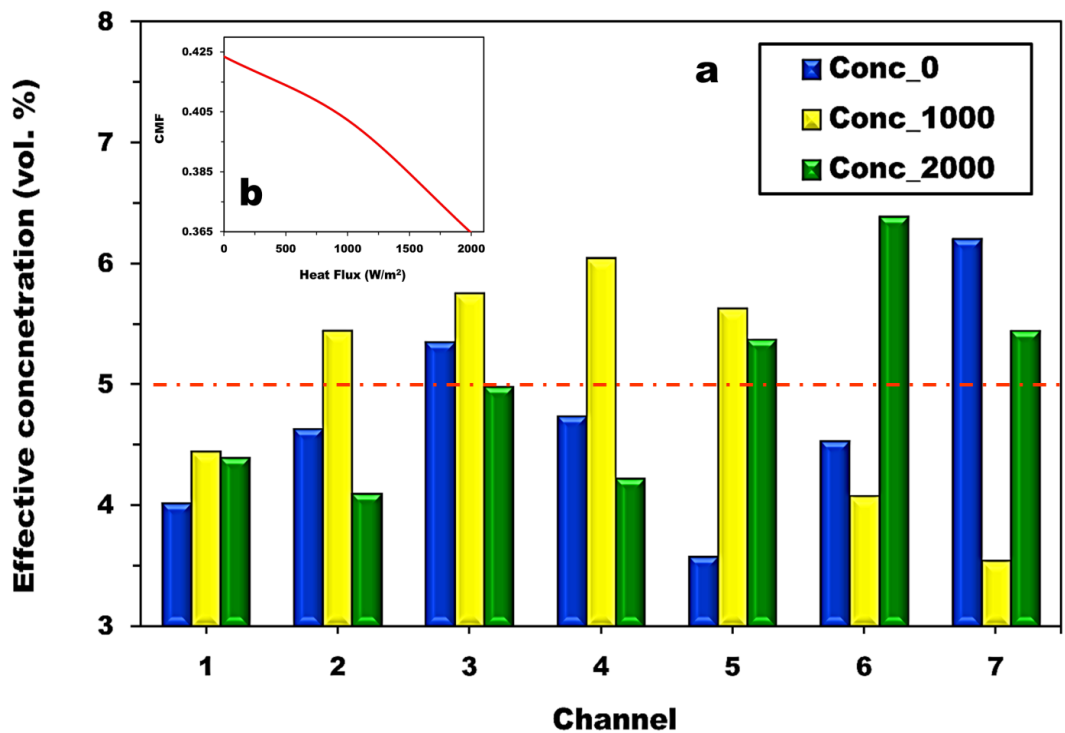

**Figure 11:** **(a)** Effective concentration in individual microchannel for different heat fluxes for Re=5 and concentrated nanofluid (5 vol. %). **(b)** The behavior of the CMF at different heat fluxes for conditions equivalent to (a).

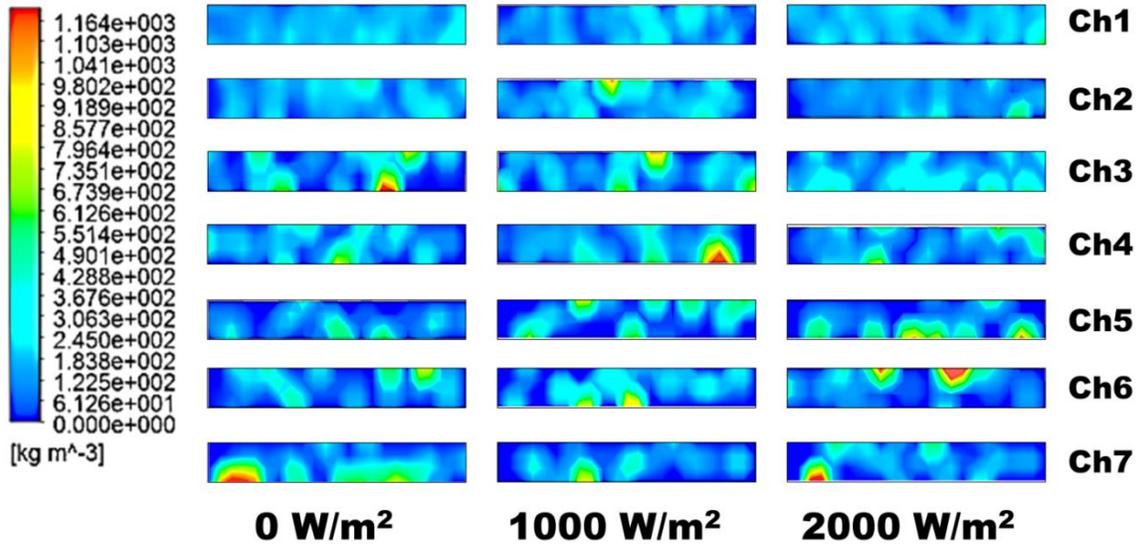

**Figure 12:** Particle mass concentration distribution contours in individual channels at different heat fluxes for Re=5 and 5 vol. %.

Having discussed the distribution patterns of the nanoparticles in the individual channels and also the effect of temeprature on the same, more insight can be shed onto the subject matter by considereing the effect of temperature on the distribution cross section of the nanoparticles along a particular channel. Fig. 13 illustrates the concentration cross section for the nanoparticles within channel 3 (for adiabatic conditions) and channel 5 (at 2 kW/m$^2$). These cases have been meticulously chosen as these channels receive nanofluid flow of the same effective concentration (evident from Fig. 11). As observable, the distribution pattern in case of the adiabatic flow remains qualitativley similar whereas that of the diabatic case clearly shows signs of redistribution and mixing. The fact that the particles tend to stay away from the heated bottom wall of the channel in the diabatic case further proves the vital effect of thermophoresis in particle distribution and subsequent heat transport. As the flow traverses towards the end of the channel, it gathers more heat and the Brownian flux increases, leading to more diffused distribution than that of the channel entrance regions.

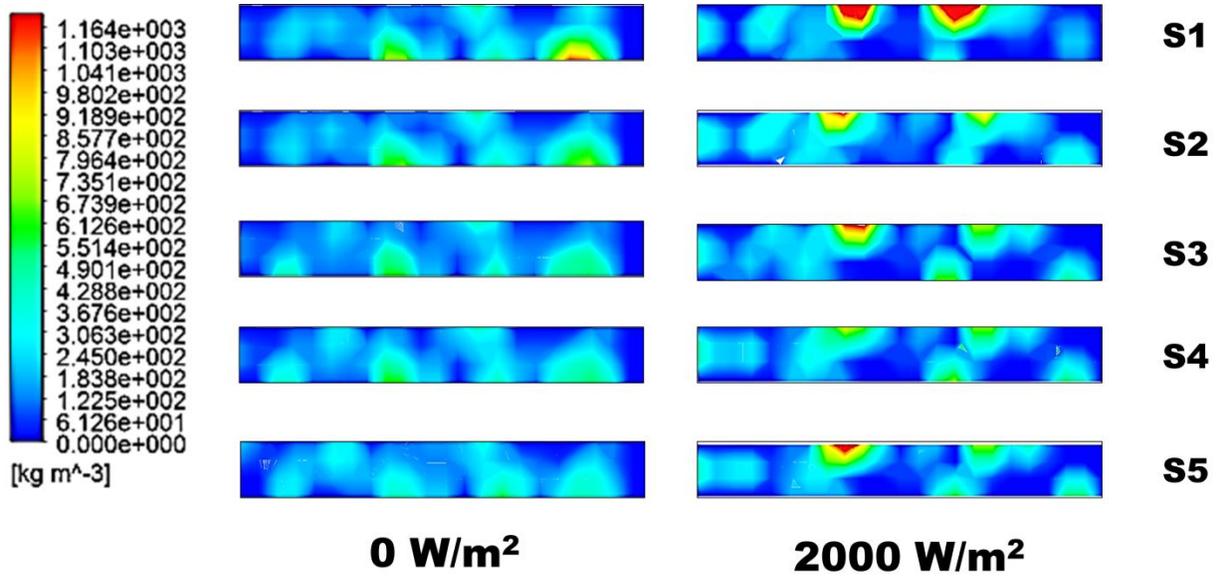

**Figure 13:** Particle mass concentration contours of nanoparticles at Re = 5 and 5 vol. % nanofluid in channel 3 and channel 5 at different cross sections for 0 W/m$^2$ and 2 kW/m$^2$ respectively. The channels have been so chosen since they have the same equivalent concentrations entering from the manifold (Fig. 11 (a)).

### 3.2.3. Thermal performance

The cooling capability of the nanofluids is of course a major focus of the present article. Fig. 14 (a) and (b) illustrate the difference between the temperatures at the inlet and outlet in individual channels at 1000 W/m$^2$. The temperature drop different Re for 5 vol. % nanofluid has been shown in Fig. 10 (a). From the figure it can be inferred that the temperature drop for initial channels is less when compared to those of end channels and this is because of non–uniform distribution of fluid due to flow maldistribution. Increase in Re enhances the heat transfer coefficient as well as the increased inertia leads to better fluid distribution among later channels, leading to less temperature drop across the channels. As discussed earlier, the presence of nanoparticles in base fluid leads to change in flow distribution among channels and this effect is more obvious at low Re and high concentrations (as illustrated in Fig. 3(b)) and the same can also be observed in Fig 10 (a) and (b). At high Re, the temperature drop in channels follow an inclined line with a constant slope whereas the slope of the line gradually increases at low Re. At low inertia flow regimes (Re=5), the temperature drop is higher towards the end channels which is caused by higher degree of maldistribution of fluid at low Re and high concentrations. The temperature drop in channels at different concentrations for Re=5 has been shown in Fig. 10 (b) and it clearly demonstrates the effectiveness of nanofluids of conventional fluids.

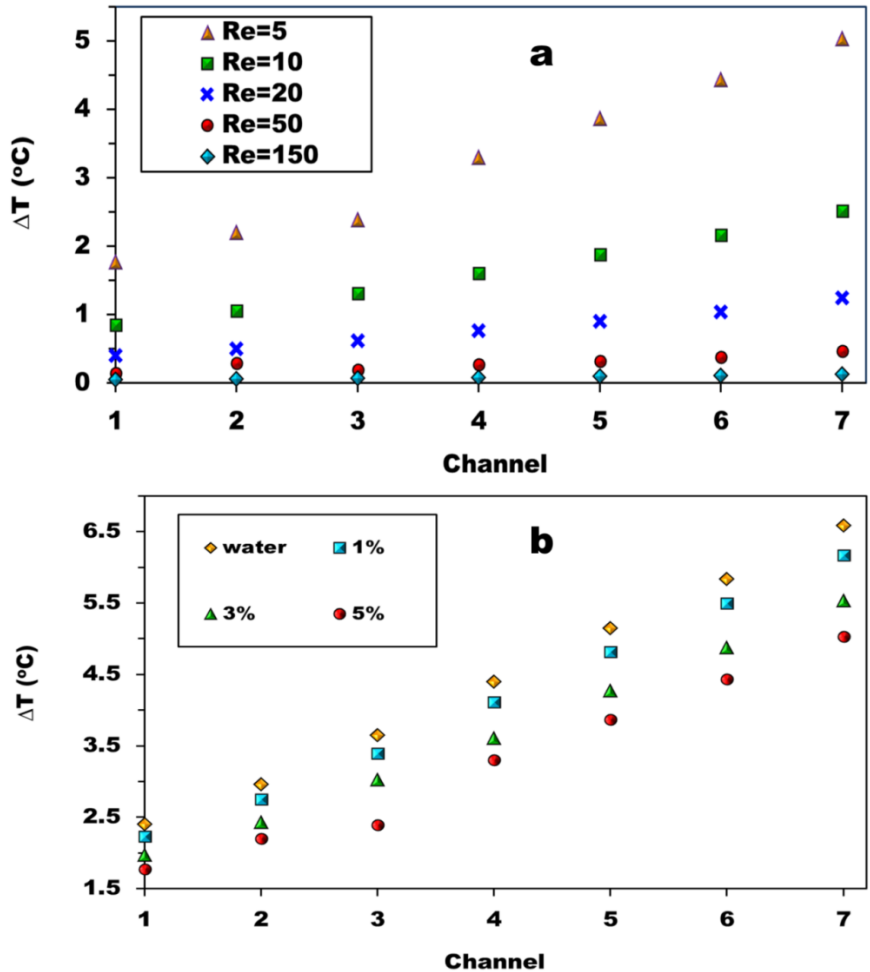

**Figure 14:** Temperature difference between the inlet and the outlet of the channels at 1 kW/m$^2$ (a) Temperature difference at different Re for 5 vol. % nanofluid (b) Temperature difference at Re = 5 for water and different concentrations of nanofluid.

As with the case of flow, the efficacy of the EPM and DPM in predicting heat transfer by nanofluids in microchannel systems also requires further probing. For this purpose, maximum temperature occurring within the flow domain are considered and illustrated in fig. 15 (a). It can be observed that the results obtained from EPM analysis are consistently higher than those obtained from DPM analysis and this difference increases with increase in heat load. Essentially, the figure further shed light onto the effectiveness of DPM in modeling convective transport in nanofluids. Since the maximum temperature within the domain is lower in case of DPM than EPM, it essentially means that the non–homogeneous nature of the nanofluids lead to efficient cooling of hotspots within the domain, thus establishing the '*smart* fluid' characteristics of nanofluids and the efficacy of DPM in capturing the same. The EPM predicts higher degrees of thermal maldistribution compared to DPM, up to 5 $^o$C in case of 5 kW/m$^2$ heat load and Re =5, since it does not account the particle migration effects. Effects such as enhanced Brownian and thermophoretic flux due to high heat flux leads to enhanced transport of heat from the channel walls to the bulk fluid as well as modifies particle distribution patterns (as discussed earlier), leading to more cooling, both in magnitude and uniformity. It is only beyond Re values of 50 that the predictions by EPM are similar to that of DPM since at high velocity flows, the migration effects are arrested and the cooling essentially occurs due to increased mass flux of fluid. However, for microscale devices where low Re flows are expected in reality, such analysis is required for predicting cooling capabilities of nanofluids as working fluids. A larger magnitude of the standard deviation of temperatures at a statistical population of data points in the domain essentially signifies more non–uniformity in the cooling characteristics. The differences between the standard deviations obtained for EPM and DPM based computations have been illustrated in Fig. 15 (b). As observable from Fig. 15 (b), nanofluids are more effective cooling fluids than water, irrespective of the model employed for prediction. The caliber of the EPM can be seen to deteriorate with increasing concentration and this is further evidence that the particle migration and diffusive events are major governing parameters towards understanding thermofluidic performance of nanofluids.

Finally, having established the physics of flow distribution of nanofluids in parallel microchannel systems and the overall cooling effectiveness, it deems a necessity to mathematically predict the cooling capability of a given nanofluid for a particular geometry so as to reduce experimental trials for system optimization. The performance of a fluid in cooling a complex geometry can be assessed from the average temperature of the system and the standard deviation of a statistical population of temperatures. It has already been established that nanofluids are better coolants when the average temperature is concerned, as it is lower than that due to the base fluid itself. Furthermore, as shown in Fig. 15 (b), the standard deviation of the temperatures of a large number of points in the heated domain is also low in case of nanofluids, thereby proving that these fluids not only cool a system better but does the same much more uniformly than normal fluids. However, the extent of this uniformity needs to be mathematically predicted in order to understand the effects of nanofluid concentration and flow domains on the cooling caliber. From analysis of data, the standard deviation of the temperature drop in the channels (proposed here as the Cooling Performance Uniformity Factor) can be related to standard deviation for water as base fluid (at same Re), the Re and concentration are expressed as points extracted from the total domain, for different Re. The high cooling capability of nanofluids can solely be predicted by the DPM which incorporates particle migration effects. **(b)** The increasing performance of DPM in prediction of cooling caliber of nanofluids with increasing concentration validates the importance of particle migration effects in such fluids.

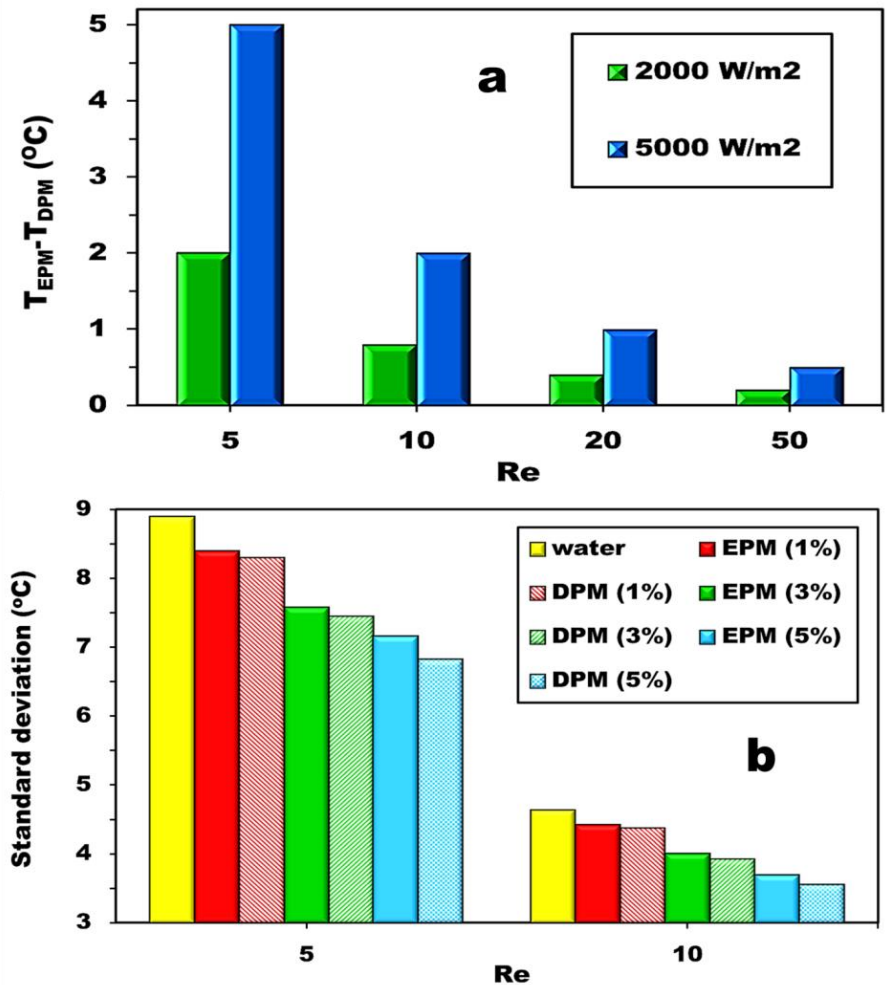

**Figure 15:** Quantitative illustration of the efficiency of the DPM over the conventional EPM in prediction thermofluidics features of nanofluid flows. **(a)** Difference between the maximum temperatures, as obtained from EPM and DPM simulations, of a multitude of temperature data

$$CPUF = \sigma(\Delta T_{ch,nf})\big|_{Re} = \left\{\sigma(\Delta T_{ch,bf})\big|_{Re} - \left(\frac{Re_{crit}}{Re}\right)\phi\right\} \tag{24}$$

The predictions obtained from the described equation have been compared with respect to the predictions obtained from full scale simulations and the same have been illustrated in Fig. 16.

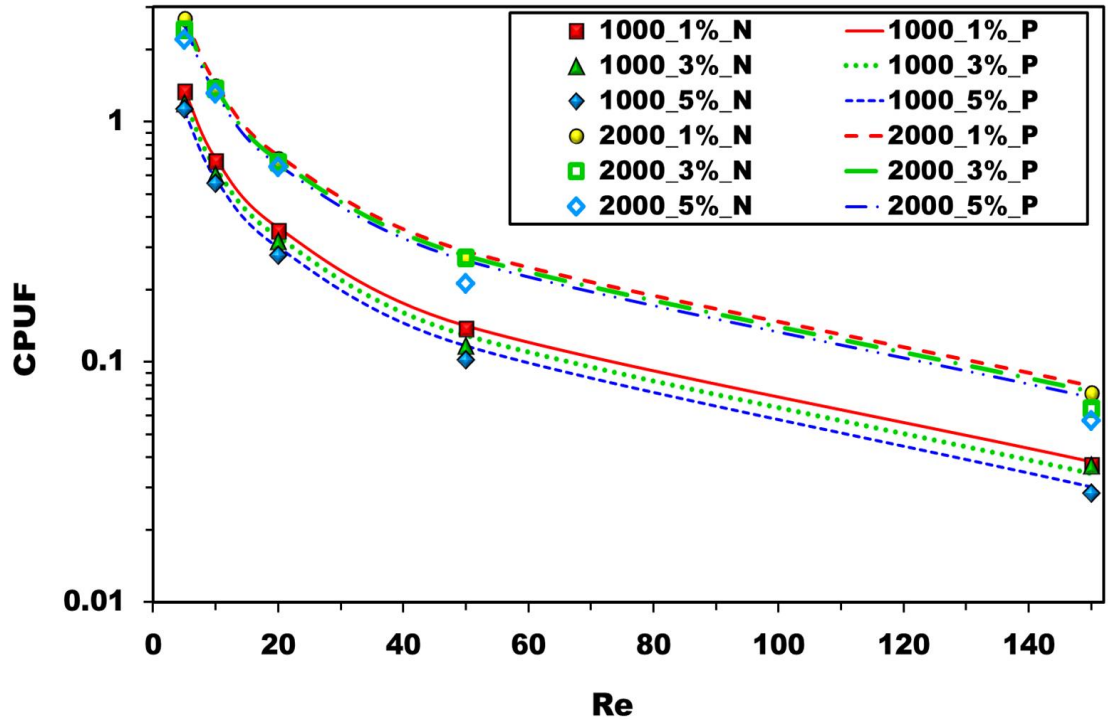

**Figure 16:** Cooling performance prediction for nanofluids in microchannel heat exchanger systems.

The uniformity parameter has been found to be a direct function of concentration, i.e., the reduction in standard deviation compared to that of water is higher when concentrated nanofluids are employed. While this sounds promising, very high concentrations lead to excessive pumping power and stability issues for the nanofluid in reality. Accordingly, the concentration requires being optimal so as to obtain minimal increment in pumping power and maximum possible uniformity in cooling. Similarly, the inverse relation to the Re implies that flows of higher velocity lead to reduction in uniformity and this has been observed before that as Re increases, the behavior tends towards that of a homogeneous fluid. Accordingly, the Re also requires optimization so as to obtain maximal uniformity in cooling but should not be too low such that the average cooling performance deteriorates at the expense of uniformity. The effect of geometry comes into the picture through the critical Re value, which is purely dependent on the geometry. At low Re, the CMF increases with increment in concentration and at higher Re, it decreases. The transit Re value at which the CMF becomes independent of the concentration is termed as the critical Re and can be deduced from analysis of simulation results. For the present geometry, this is determined to be ~ 30 and a clear scrutiny of Fig. 5 shows that the CMF is fairly constant for Re = 50, thus providing credibility to the obtained value.

## 4. Conclusions

To infer, the present article deals with the flow and concentration maldistribution of nanofluids in parallel microchannel systems. Reports in literature treat nanofluids as homogeneous single phase fluids with enhanced effective properties and conclude improved cooling performance in such devices. However, experiments reveal that such predictions fall short of the real fluid distribution and cooling performance and hence a non–homogeneous two phase model has to be utilized to model nanofluid flows. In this article, an Eulerian–Lagrangian model for nanofluid flow in U configuration parallel microchannels has been considered and distribution of particles as well as the fluid and their impact vis-à-vis thermal performance has been reported. It has been observed that EPM cannot be used to predict nanofluid performance in complex flow geometries as the distribution of particles and the fluid are inter–dependent on the distribution patter of one another. This leads to grossly different flow distribution patterns and the effective particle concentration flowing in the individual channels. This distribution is further dependent on temperature and the distribution has been observed to be more uniform at high heat fluxes. Essentially, this leads to *'smart'* and more uniform cooling and this is only predicted by DPM formulation. A mathematical predictive model has also been proposed to determine a quantitative measure of the uniformity of cooling performance of the nanofluid over water as base fluid. The present findings can be utilized to obtain *a priori* estimates of nanofluid behavior within a particular micro–geometry for optimizing flow and thermal performance in parallel microchannel heat sinks employing nanofluid coolants.

## Acknowledgements

The authors thank the Defence Research and Development Organization (DRDO) of India for partial financial support for the computational facilities. (Grant no. ERIP/ER/RIC/2013/M/01/2194/D (R&D)). LSM would like to thank the Ministry of Human Resource Development (Govt. of India) for the doctoral scholarship. PD would like to thank IIT Madras for the post–doctoral fellowship.